\documentclass{aa}
\usepackage{graphicx}
\usepackage{epsfig}
\usepackage{graphics}
\usepackage{subfigure}
\usepackage{natbib}
\bibpunct{(}{)}{;}{a}{}{,} 

\begin{document}

\title{Non-thermal processes around accreting galactic black holes}

\author{G. E. Romero\inst{1,2,}\thanks{Member of CONICET, Argentina}, F. L. Vieyro\inst{1,}\thanks{Fellow of CONICET, Argentina} \and G. S. Vila\inst{1,\star \star}}
  
\institute{Instituto Argentino de Radioastronom\'{\i}a (IAR, CCT La Plata, CONICET), C.C.5, (1984) Villa Elisa, Buenos Aires, Argentina \and Facultad de Ciencias Astron\'omicas y Geof\'{\i}sicas, Universidad Nacional de La Plata, Paseo del Bosque s/n, 1900, La Plata, Argentina}

\offprints{F. L. Vieyro \\ \email{fvieyro@iar-conicet.gov.ar}}

\titlerunning{Non-thermal processes around black holes}

\authorrunning{G.E. Romero, et al.}

\abstract
{Accreting black holes in galactic X-ray sources are surrounded by hot plasma. The innermost part of these systems is likely a corona with different temperatures for ions and electrons. In the so-called low-hard state, hot electrons Comptonize soft X-ray photons from the disk that partially penetrates the corona, producing emission up to $\sim 150$ keV, well beyond the expectations for an optically thick disk of maximum temperature $\sim 10^{7}$ K. However, sources such as Cygnus X-1 produce steady emission up to a few MeV, which is indicative of a non-thermal contribution to the spectral energy distribution.} 
{We study the radiative output produced by the injection of non-thermal (both electron and proton) particles in a magnetized corona around a black hole.} 
{Energy losses and maximum energies are estimated for all types of particles in a variety of models, characterized by different kinds of advection and relativistic proton content. Transport equations are solved for primary and secondary particles, and spectral energy distributions are determined and corrected by internal absorption. } 
{We show that a local injection of non-thermal particles can account for the high energy excess observed in some sources, and we predict the existence of a high-energy bump at energies above 1 TeV, and typical luminosities of $\sim 10^{33}$ erg s$^{-1}$.} 
{High-energy instruments such as the future Cherenkov Telescope Array (CTA) can be used to probe the relativistic particle content of the coronae around galactic black holes.} 
 
 
\maketitle

\section{Introduction}

Disks around galactic black holes are formed by the accretion of matter with angular momentum. These disks are opaque and emit locally thermal radiation of temperatures up to $\sim10^{7}$ K \citep{shakura}. These disks cannot be responsible for the hard X-ray emission extending up to a few MeV detected from Cygnus X-1 and similar X-ray binaries and microquasars in the low-hard state. This finding and the presence in the spectrum of features like a broad iron K$\alpha$ line and a hard X-ray bump, led to the idea that a corona of hot plasma might surround the black hole and part of the disk. Two-temperature models for this plasma were first suggested by \citet{shapiro}. In these types of models protons are much hotter than electrons ($ T_{i}>>T_{e}$). Since the pressure is dominated by the protons, the disk inflates, the density drops, and there is a low rate of Coulomb energy exchange between protons and electrons, allowing the existence of a two-temperature plasma in a self-consistent way. The Comptonization of soft photons results in a hard X-ray spectrum. The model, however, is unstable to small perturbations in the ion temperature $T_{i}$ \citep{pringle,piran}.

When the plasma density is very low, the protons are unable to transfer energy to the electrons. If matter is advected into the black hole \citep{ichimaru,narayana,narayanb} or removed outward as a hot wind \citep{blandford01}, thermally stable solutions can be found. The geometry of the region with the hot plasma is not well-constrained, but a spherical region around the black hole is usually considered \citep[e.g.,][]{esin01,esin02}. The two-temperature plasma then forms a hot corona around the black hole. In the different spectral states, the cold disk penetrates to different distances from the black hole. In the very high-soft state it goes all the way down to the last stable orbit (see \citealt{narayan02,narayan03}, for comprehensive reviews).

The general view of a relatively low density hot corona and a cold accretion disk was presented by \citet{bisnovatyi}. The mechanism for heating the corona may be magnetic reconnection of field loops emerging from the disk \citep{galeev}. Violent reconnection may lead to plasma motions and collisions, with shock formation. A non-thermal particle population might then arise in the corona as the result of diffusive shock acceleration \citep[e.g.,][]{spruit}. 

The effects of a non-thermal population of electrons in a hot corona were considered by \citet{kusunose} and more recently by \citet{belmont} and \citet{vurm}. The results of the injection of non-thermal protons and secondary pions and muons in a magnetized corona has not been comprehensively studied so far. The contributions from the transient particles can be important at high energies. The emerging emission from all non-thermal processes may in principle be detectable by future Cherenkov telescope arrays such as CTA or AGIS. Hence, high-energy gamma-ray astronomy can provide a tool to probe the non-thermal particle content of hot coronae around black holes. 

In this paper, we present detailed calculations of the radiative output of non-thermal particles in a simplified model of magnetized corona. The existence of the corona is assumed and the effect of injection of both relativistic protons and electrons are considered. The distributions of all relevant secondary particles are estimated and their radiative output computed. The coronal matter and radiative fields are considered as targets for the populations of relativistic particles. The internal absorption of gamma-rays is calculated and the spectral energy distributions of different models are presented. The result is a self-consistent treatment of the non-thermal processes in the sense that once the medium is fixed in each model, we solve the transport equations for all type of particles and use the obtained particle distributions to estimate the non-thermal radiation.

The structure of the paper is as follows. In Sect. \ref{scenario}, we outline the basic scenario that is discussed in the paper. Section \ref{losses} deals with particle acceleration and losses in the coronal environment. The maximum energy for the different particles is determined. In Sect. \ref{SEDs}, the radiation is calculated and the spectral energy distributions are presented, for different sets of parameters. Section \ref{Cyg} presents an application to Cygnus X-1. We close with a brief discussion and conclusions in Sects. \ref{Discussion} and \ref{Conclusions}, respectively.

\section{Basic scenario}\label{scenario}

\begin{figure*}[!t]
\centering
\includegraphics[clip,width=12 cm]{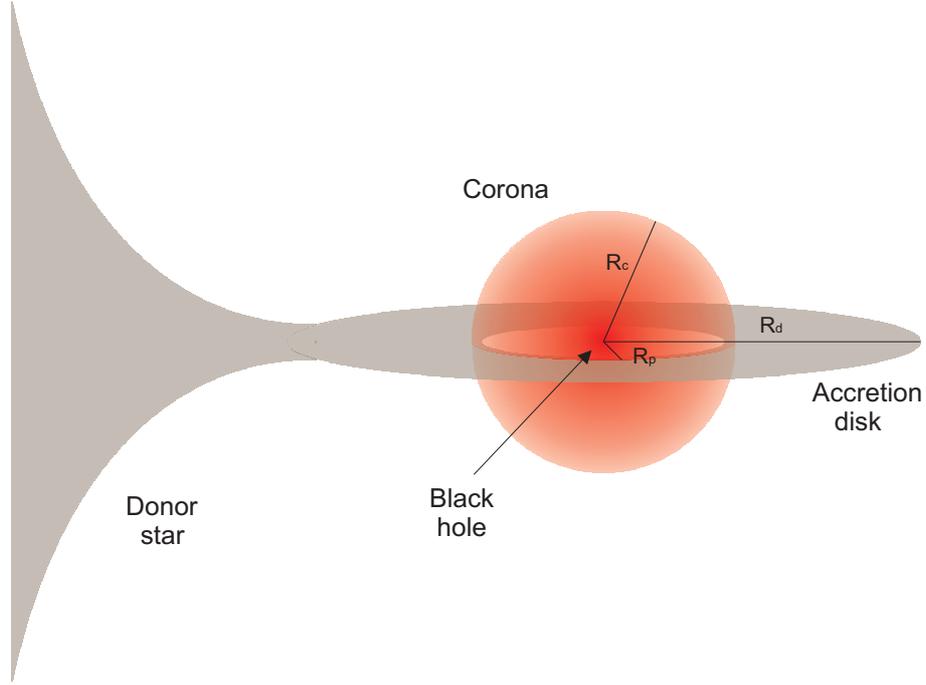}
\caption{Schematic representation of the components of the system discussed in the text. In the spherical corona, the thermal and non-thermal components are co-spatial. Not to scale.}
\label{fig:geometria}
\end{figure*}

The low-hard state of accreting black holes is characterized by the presence of a hot corona around the compact object. The existence of this component is strongly implied by the X-ray spectrum of Cygnus X-1 \citep[see][]{dove,esin02}. Figure \ref{fig:geometria} shows a sketch of the main components of this system. We assume a spherical corona with a radius $R_{\rm{c}}$ and an accretion disk that penetrates the corona up to $R_{\rm{p}}<R_{\rm{c}}$. For simplicity, we consider the corona to be homogeneous and in a steady state.

In our model, we assume that the luminosity of the corona is 1 \% of the Eddington luminosity of a 10 $M_{\odot}$ black hole. In Table \ref{table}, we present the values of the main parameters adopted and derived for the system (they are the typical values of the hard X-ray luminosity of Cygnus X-1, see \citealt{poutanen01}). The resulting corona has a luminosity $L_{\rm{c}}=1.3 \times 10^{37}$ erg s$^{-1}$, in accordance with observations.

X-ray binaries with well-formed coronae in the low-hard state produce relativistic jets. In these systems the power of the jet is understood to be related to the magnetic field, since the jet launching mechanism is likely of magnetic origin \citep[e.g.,][]{spruit}. For Cygnus X-1 and similar systems, the jet kinetic power is of approximately the same luminosity as the corona. Hence, the value of the mean magnetic field $B$ can be estimated by equipartition between the magnetic energy density and the bolometric photon density of the corona $L_{\rm{c}}$ (e.g., \citealt{bednarek},
and references therein).
	
	\begin{equation}
		\frac{B^2}{8\pi}=\frac{L_{\rm{c}}}{4\pi R_{\rm{c}}^2c} ,
	\end{equation}

\noindent which implies a value $B=5.7\times 10^5$ G.

It is assumed that the corona consists of a two-temperature plasma, with an electron temperature $T_{e}=10^{9}$ K and an ion temperature $T_{i}=10^{12}$ K \citep[e.g.,][]{narayana,narayanb}. The physical conditions in which this assumption is valid were extensively discussed by \citet{narayana,narayanb} and \citet{narayan02}. The corona and the base of the jet are thought to be a region with similar properties in the standard jet-disk symbiosis model \citep{malzac}. We assume a spherical corona where thermal and non-thermal components are co-spatial. If the magnetic field launches the plasma into a jet, equipartition between magnetic and kinetic energy densities is a reasonable assumption that allows us to estimate the plasma density \citep[e.g.,][]{zdziarski,narayana,narayanb} to be
			
			\begin{equation}
				\frac{B^2}{8\pi}=\frac{3}{2}n_{e}kT_{e}+\frac{3}{2}n_{i}kT_{i} ,
			\end{equation}

\noindent where $n_{i}$ and $n_{e}$ are the ion and electron densities, respectively. For a corona mainly composed of hydrogen, this implies that $n_{i}\sim n_{e}=6.2\times 10^{13}$ cm$^{-3}$.

The hard X-ray emission of the corona is characterized by a powerlaw with an exponential cutoff at high energies, as observed in several X-ray binaries in the low-hard state \citep[e.g.,][]{romero01} for which

			\begin{equation}
				n_{\rm{ph}}(E)=A_{\rm{ph}} E^{-\alpha}  e^{-E/E_{\rm{c}}}\; \textrm{erg$^{-1}$ cm$^{-3}$.}
			\end{equation}

\noindent In accordance with the well-studied case of Cygnus X-1 \citep[e.g.,][]{poutanen01}, we adopt $\alpha=1.6$ and $E_{\rm{c}}=150$ keV. The normalization constant $A_{\rm{ph}}$ can be obtained from $L_{\rm{c}}$,
	
			\begin{equation}
				\frac{L_{\rm{c}}}{4\pi R_{\rm{c}}^2c} = \int_{0}^\infty {E n_{\rm{ph}}(E)dE} = \int_{0}^\infty {A_{\rm{ph}} E^{1-\alpha}e^{-E/ E_{\rm{c}}}dE } \textrm{.}
			\end{equation}

\begin{table}[!t]
    \caption[]{Model parameters}
\begin{tabular}{ll}
\hline\noalign{\smallskip}
Parameter & Value\\[0.01cm]
\hline\noalign{\smallskip}
$M_{\rm{BH}}$:  black hole mass [$M_{\odot}$]							& $10$$^{(1)}$					\\[0.01cm]
$R_{\rm{c}}$:   corona radius [cm] 										 	  & $5.2 \times 10^{7}$$^{(1,2)}$ 	\\[0.01cm]
$\bf{R_{\rm p}/R_{\rm{c}}}$:   ratio of inner disk to corona radius  										 	  & $0.9$$^{(1)}$ 	\\[0.01cm]
$D(1+S)$:       Covering fraction of the corona           & $0.08$$^{(1)}$        	\\[0.01cm]
around thermal disk times the seed                        &                         \\[0.01cm]
photon correction factor                                  &                         \\[0.01cm]
$T_{e}$:        electron temperature [K] 								&	$10^9$								\\[0.01cm]
$T_{i}$:        ion temperature [K] 											&	$10^{12}$ 						\\[0.01cm]
$E_{\rm{c}}$:   X-ray spectrum cutoff [keV]							& $150$       					\\[0.01cm]
$\alpha$: 			X-ray spectrum power-law index    				& $1.6$									\\[0.01cm]
$\eta$: 				acceleration efficiency 									& $10^{-2}$							\\[0.01cm]
$B_{\rm{c}}$: 	magnetic field [G]				 								& $5.7 \times 10^5$				\\[0.01cm]
$n_{i},n{e}$:   plasma density [cm$^{-3}$] 								& $6.2 \times 10^{13}$	\\[0.01cm]
$A_{\rm{ph}}$:  normalization constant [erg$^{3/5}$ cm$^{-3}$] 
																										& $2.6 \times 10^{12}$ 				\\[0.01cm]
$a$: 						hadron-to-lepton energy ratio 						& $1$ - $100$						\\[0.01cm]
$kT$:						disk characteristic temperature [keV] 	  & $0.1$									\\[0.01cm]
$v$:						advection velocity  [c] 							 		& $0.1$									\\[0.01cm]

\hline\\[0.005cm]
\multicolumn{2}{l}{
$^{(1)}$ Typical value for Cygnus X-1 in the low-hard state }				  \\[0.01cm]
\citep{poutanen01}. \\[0.01cm]
\multicolumn{2}{l}{
$^{(2)}$ $35 R_{\rm{G}}$, $ R_{\rm{G}}=\frac{GM}{c^{2}}$.} 			 																				\\[0.01cm]
\end{tabular}	
  \label{table}
\end{table}

\section{Particle acceleration and losses}\label{losses}

We now consider the interaction of locally injected relativistic particles with the matter, photons, and magnetic fields of the corona and the disk, which are taken as background components.

There are three relevant processes of interaction of relativistic electrons and muons with these fields: synchrotron radiation, inverse Compton (IC) scattering, and relativistic Bremsstrahlung. For protons and charged pions, there are also three important processes: synchrotron radiation, proton-proton (or pion-proton) inelastic collisions, and photohadronic interactions. The neutral pions have a short mean lifetime of $8.4 \times 10^{-17}$ s, and they therefore decay before interacting.

We also estimated the energy losses caused by magnetic Bremsstrahlung. This contribution mainly falls at intermediate energies. Since this energy range is completely dominated by the thermal emission from the star and the accretion disk, we do not show the magnetic Bremsstrahlung luminosity in the SEDs shown below.

The synchrotron cooling rate for a particle of mass $m$, charge $e$, and energy $E$ in a region of magnetic energy density $U_B$ is

\begin{equation}\label{eq:coolsyn}
	t^{-1}_{\rm{synchr}}=\frac{4}{3}\left(\frac{m_e}{m}\right)^3\frac{c\sigma_TU_B}{m_ec^2}\frac{E}{mc^2}.
\end{equation}  

\noindent In both Thomson and Klein-Nishina regimes, the IC cooling rate for an electron is given by \citep{blumenthal}
		
		\begin{equation}				t^{-1}_{\rm{IC}}=\frac{1}{E_{e}}\int_{\epsilon_{\rm{min}}}^{\epsilon_{\rm{max}}}\int_{\epsilon}^{\frac{\Gamma E_{e}}{1+\Gamma}}{(\epsilon _{1}-\epsilon)\frac{dN}{dtd\epsilon_{1}}d\epsilon_{1}} .
			\end{equation}
			
\noindent Here $\epsilon$ is the energy of the incident photon, $\epsilon _{1}$ is the energy of the scattered photon, and  

		\begin{equation}
			\frac{dN}{dtd\epsilon_{1}}  =  \frac{2\pi r_{0}^2 m_{e}^2 c^5}{E_{e}^2} \frac{n_{\rm{ph}}(\epsilon) d \epsilon}{\epsilon} F(q),
		\end{equation}
		
\noindent where	$n_{\rm{ph}}(\epsilon)$ is the density of target photons, $r_{0}$ the classical radius of the electron, and
		
		\begin{eqnarray}\label{IC}
			  F(q) 	&=& 2q\ln{q}+(1+2q)(1-q)+\frac{1}{2} (1-q) \frac{(\Gamma q)^2}{1+\Gamma q} ,  \nonumber\\
			  \Gamma 		&=& 4\epsilon E_{e} / (m_{e}c^2)^2,   \\
			  q 		&=& \epsilon_{1}/[\Gamma (E_{e}-\epsilon_{1})]  \nonumber.
		\end{eqnarray}

  We consider two target photon fields: the power-law photon field of the corona and the field from the disk. The latter can be represented by a black body with temperature $kT=0.1$ keV \citep{poutanen01}. The radiation field in the corona is diluted to account for the solid angle subtended by the disk as seen from the corona. This is performed by means of a parameter $D$ that indicates the fraction of the light emitted by the thermal region, that passes through the corona, and the parameter $S$ that is the ratio of intrinsic seed photon production in the corona to the seed photon luminosity injected from outside. Since we apply our model to Cygnus X-1, we follow the estimates of \citet{poutanen01} for this source that are based on the analytic scaling approximations for the thermal Comptonization spectrum given by \citet{pietrini}. We consider the value of the different parameters that are relevant to the low-hard state. In particular, $D(1+S)=0.08$, and the ratio of the disk inner radius to the corona radius is taken to be $\approx 0.9$. For additional details, the reader is referred to \citet{poutanen01}. For a general picture of the Comptonization process, \citet{done} is a very useful reference.    

For a completely ionized plasma, the Bremsstrahlung cooling rate is \citep[e.g.,][]{berezinskii}
			
			\begin{equation}
				t^{-1}_{\rm{Br}}=4n_{i}Z^2r_{0}^2\alpha c\left[\ln \left(\frac{2E_e}{m_{e}c^2}\right)-\frac{1}{3} \right] .
			\end{equation}

\noindent The cooling rate for inelastic collisions of protons with nuclei of the corona is given by:

			\begin{equation}\label{eq:pp}
				t^{-1}_{pp}=n_{i}c\sigma _{pp}K_{pp},
			\end{equation}
			
\noindent where $K_{pp}$ is the total inelasticity of the process, of value $\sim 0.5$. The total cross-section $\sigma_{pp}$ can be approximated by \citep{kelner}
	
			\begin{equation}
				\sigma_{\rm{inel}}(E_{p})=(34.3+1.88L+0.25L^2)\left[ 1-\left( \frac{E_{\rm{th}}}{E_{p}}\right)^4 \right ]^2 , \label{sigma_pp}
			\end{equation}

\noindent where 
	
			\begin{equation}
				L= \ln{\left(\frac{E_{p}}{1\textrm{ TeV}}\right )} \textrm{.}  
			\end{equation}
	
\noindent The proton threshold kinetic energy for $\pi^0$ production is $E_{\rm{th}}^{\rm{kin}} \approx 280$ MeV. 

The photomeson production takes place for photon energies above $\epsilon_{\rm{th}} \approx 145$ MeV (measured in the rest-frame of the proton). Near the threshold, a single pion is produced per interaction; at higher energies, the production of multiple pions dominates. In our model, the relevant photons come from the corona and, to a lesser extent, from the accretion disk.

The cooling rate due to photopion production for a proton of energy $E_{p}$ in an isotropic photon field of density $n_{\rm{ph}}(\epsilon)$ is given by \citep{stecker}

\begin{equation}\label{eq:pgamma}
				t^{-1}_{p\gamma}(E_{p})=\frac{m_{p}^2c^5}{2E_{p}^2}\int^{{\infty}}_{\frac{\epsilon_{\rm{th}}}{2\gamma_{p}}} 				{d\epsilon \frac{n_{\rm{ph}}(\epsilon)}{\epsilon^2}} \int^{2 \epsilon	\gamma_{p}}_{\epsilon_{\rm{th}}}{d\epsilon '\sigma_{p\gamma}(\epsilon ')K_{p\gamma}(\epsilon ')\epsilon '} ,
			\end{equation}

\noindent where $\epsilon'$ is the photon energy in the rest-frame of the proton and $K_{p\gamma}$ is the inelasticity of the interaction. \citet{atoyan} introduced a simplified approach to treat the cross-section and the inelasticity, which can be written as
	
			\begin{displaymath}\label{eq:sigma_pgamma}
				\sigma_{p\gamma}(\epsilon')\approx \left\{ \begin{array}{lrl}
				340 \; \mu\textrm{barn} & \;\;\;200 \textrm{MeV} \leq \epsilon' & \leq 500\textrm{MeV}\\
				120 \; \mu\textrm{barn}  &   \epsilon' & \geq 500\textrm{MeV} , 
				\end{array} \right.
			\end{displaymath}
	
\noindent and	
	
			\begin{displaymath}\label{eq:K_pgamma}
				K_{p\gamma}(\epsilon')\approx \left\{ \begin{array}{lrl}
				0.2 & \;\;\;200 \textrm{MeV} \leq \epsilon' & \leq 500 \textrm{MeV}\\
				0.6  &   \epsilon' & \geq 500 \textrm{MeV}. 
				\end{array} \right.	
			\end{displaymath}
			
At energies below the threshold for photomeson production, the main channel of proton-photon interaction is the direct production of electrons or positrons. The cooling rate is also given by Eq. (\ref{eq:pgamma}), and the corresponding cross-section and inelasticity. The cross-section for this channel -also known as the Bethe-Heitler cross-section- increases with the energy of the photon. Both the cross-section and inelasticity approximations in the limits of low and high energies can be found in \citet{begelman} (see Appendix A).
			
The mean lifetime of charged pions in their rest-frame is $\tau_{\pi}=2.6\times10^{-8}$ s, and then they decay into muons and neutrinos; the mean lifetime of muons is $\tau_{\mu}=2.2\times 10^{-6}$ s. They decay yielding neutrinos/anti-neutrinos and electrons/positrons. In the observer rest-frame, the decay rate is

		\begin{equation}
				t^{-1}_{\rm{dec}}= \left(\frac {\tau E}{mc^2} \right)^{-1} .\\
		\end{equation}		
			
\noindent We consider two types of corona. One is an ADAF-like corona, where matter is advected to the black hole. This model was discussed in detail for Cygnus X-1 by \citet{dove} and \citet{esin02}. In this case, particles fall onto the compact object at a mean radial velocity $v = 0.1c$, the free-fall velocity \citep{begelman}. Therefore, the convection rate is
	
	\begin{equation}\label{eq:conv}
				t_{\rm{conv}}^{-1}=\frac{v}{R_{\rm{c}}} .
		\end{equation}		

\noindent The other model considered here is a static corona (e.g., supported by magnetic fields, see \citealt{beloborodov}) where the relativistic particles can be removed by diffusion. In the Bohm regime, the diffusion coefficient is $D(E)=r_{\rm{g}}c/3$, where $r_{\rm{g}}=E/(eB)$ is the giroradius of the particle. The diffusion rate is

	\begin{equation}\label{eq:diff}
				t_{\rm{diff}}^{-1}=\frac{2D(E)}{R_{\rm{c}}^2} .
		\end{equation}

\noindent The maximum energy that a relativistic particle can attain depends on the acceleration mechanism and the different processes of energy loss. The acceleration rate $t^{-1}_{\rm{acc}}=E^{-1}dE/dt$ for a particle of energy $E$ in a magnetic field $B$, in a region where diffusive shock acceleration takes place, is given by

\begin{equation}
	t^{-1}_{\rm{acc}}=\frac{\eta ecB}{E},
	\label{accrate}
\end{equation}
   
\noindent where $\eta\leq1$ is a parameter that characterizes the efficiency of the acceleration. We fix $\eta=10^{-2}$, which describes the efficient acceleration by shocks with $v_{\rm s}\sim 0.1c$ in the Bohm regime.
			
\begin{figure*}[!ht]
\centering
\subfigure[Electron losses.]{\label{fig:perdidas:a}\includegraphics[width=0.45\textwidth,keepaspectratio]{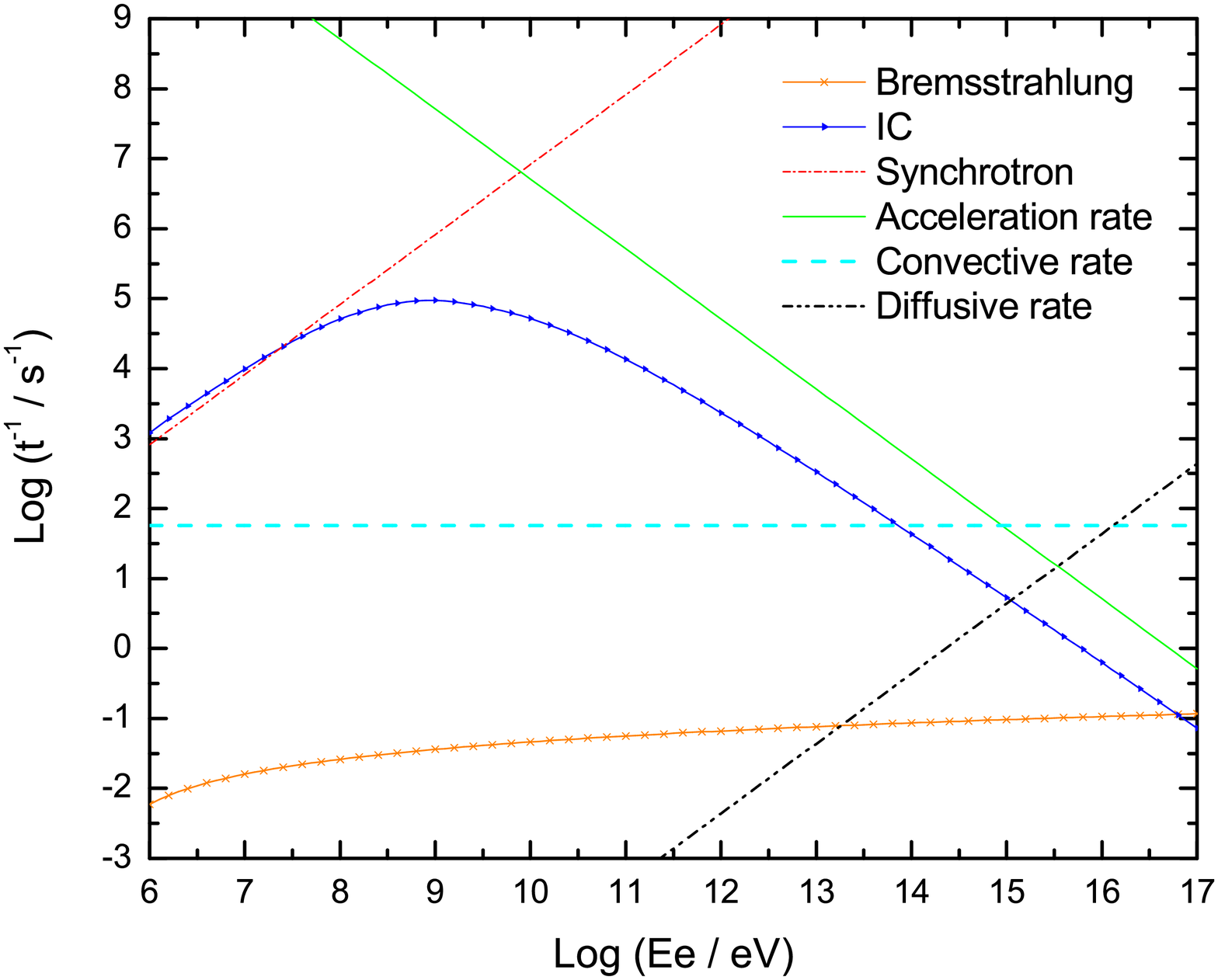}} \hspace{20pt} 
\subfigure[Proton losses.]{\label{fig:perdidas:b}\includegraphics[width=0.45\textwidth,keepaspectratio]{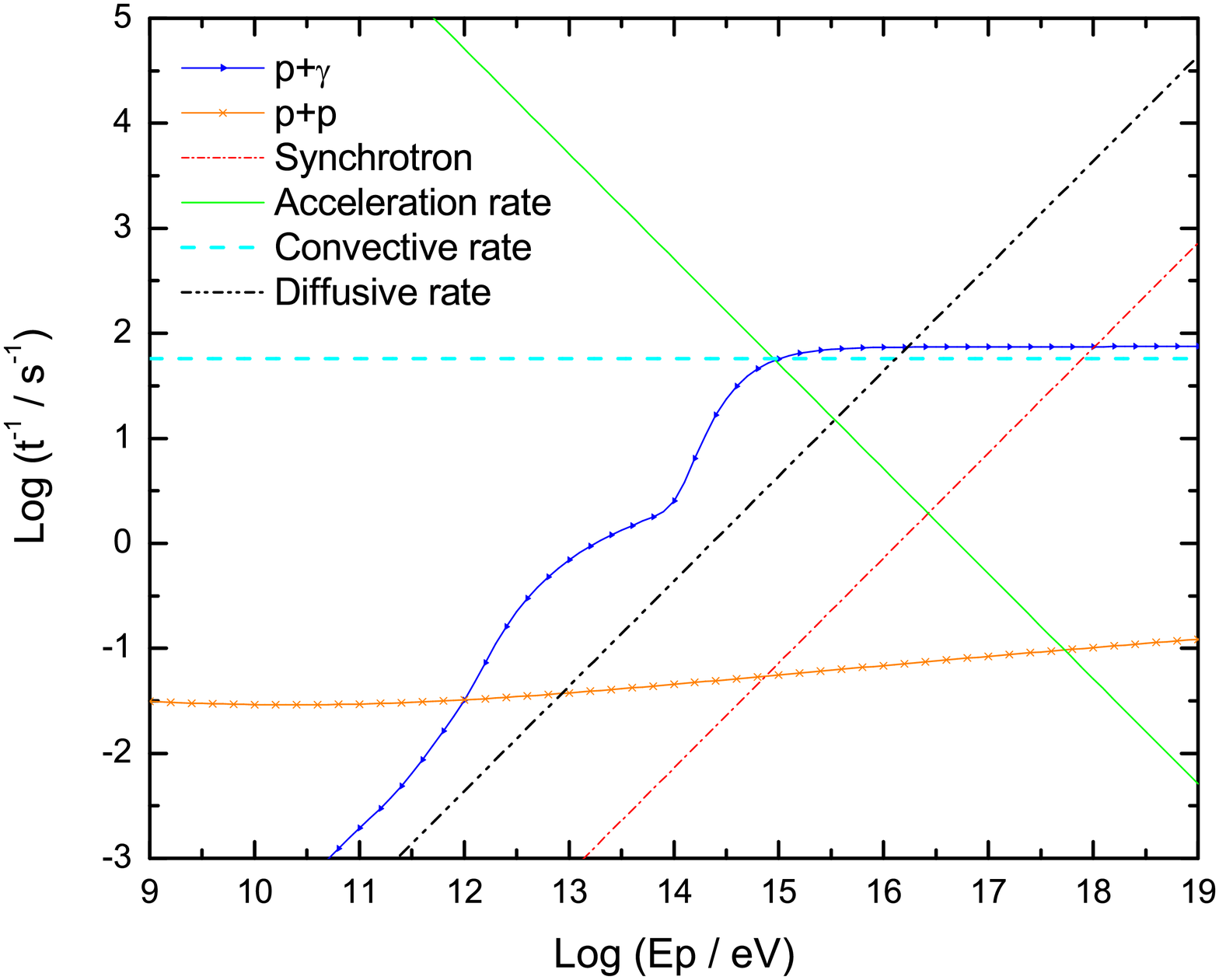}} \hfill \\ 
\subfigure[Pion losses.]{\label{fig:perdidas:c}\includegraphics[width=0.45\textwidth, keepaspectratio]{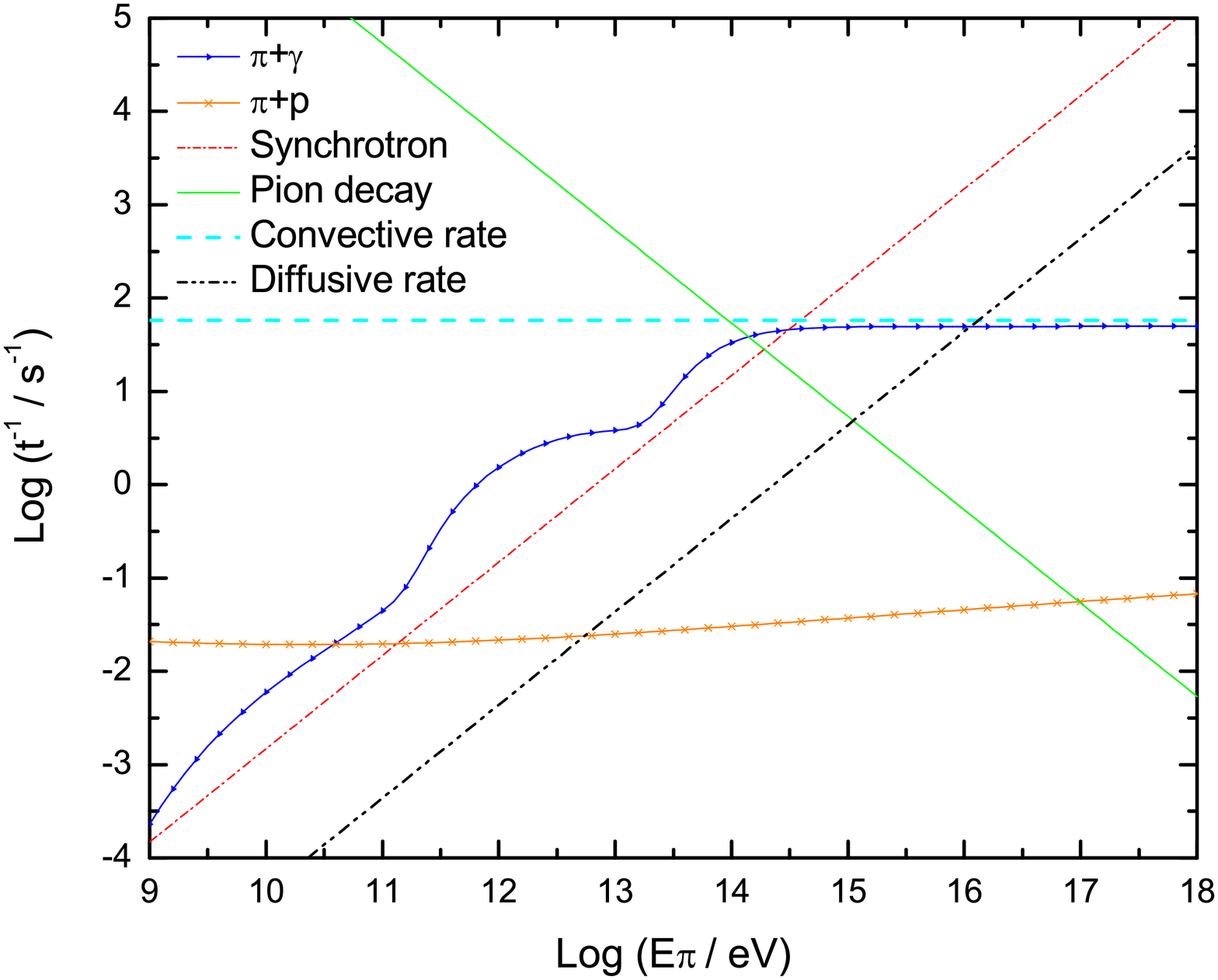}}  \hspace{20pt}
\subfigure[Muon losses.]{\label{fig:perdidas:d}\includegraphics[width=0.45\textwidth, keepaspectratio]{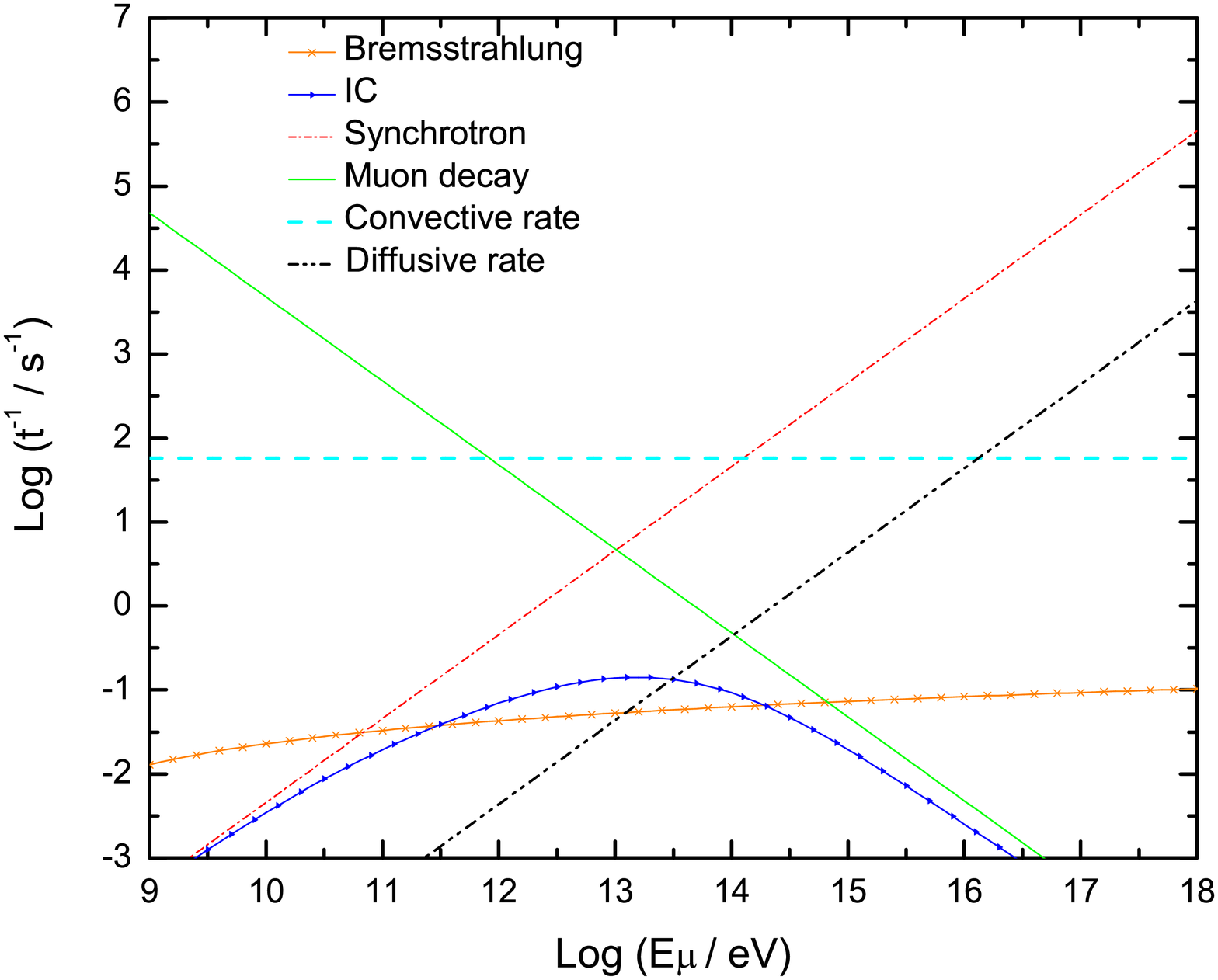}} \hfill
\caption{Radiative losses in a corona characterized by the parameters of Table \ref{table}.}
\label{fig:perdidas}
\end{figure*}

	Figure \ref{fig:perdidas} shows the cooling rates for different energy-loss processes, together with the acceleration and escape rates, for each type of particle considered. Under the physical conditions previously described, the main channel of energy loss for electrons is synchrotron radiation. Only for low-energy electrons IC losses are significant. For protons, both $pp$ and $p \gamma$ interactions are relevant.  While diffusion has almost no effect on particle distributions, advection plays a decisive role in the behavior of protons: in the model with advection, most protons fall onto the black hole before radiating their energy.			

It is possible to estimate the maximum energy achieved by the electrons equating 

\begin{equation}
			t_{\rm{acc}}^{-1}=t_{\rm{synchr}}^{-1}+t_{\rm{IC}}^{-1} ,
		\end{equation}
		
\noindent whereas for protons
		
			\begin{equation}
			t_{\rm{acc}}^{-1}=t_{pp}^{-1}+t_{p\gamma}^{-1}+t_{\rm{esc}}^{-1},
		\end{equation}

\noindent where $t_{\rm{esc}}$ is the timescale during which the relativistic particles escape from the system. This timescale is given by Eq. (\ref{eq:conv}) for models with convection, and by Eq. (\ref{eq:diff}) for models with diffusion. The maximum energies obtained by electrons and protons are $E_{\rm{max}}^{(e)}\approx 7.9\times 10^9$ eV and $E_{\rm{max}}^{(p)}\approx 8.0\times 10^{14}$ eV, respectively. These values are compatible with the Hillas criterion, given the size of the corona.

For pions, the main channel of energy loss is the $\pi \gamma$ interaction, but an important fraction of these pions decay before cooling (those of lower energies). Muons with energies above $\sim 10^{13}$ eV cool mostly by synchrotron radiation in models with a static corona. The most energetic muons fall into the black hole, in all models with dominant convection. 

The steady state particle distributions $N(E)$ can be derived from the solution to the transport equation \citep{ginzburg}

\begin{equation}\label{eq:transporte}
		\frac{\partial }{\partial E} \Big( \frac{dE}{dt} \Big | _{\rm{loss}} N(E) \Big)+ N(E)(t_{\rm{esc}}^{-1}+t_{\rm{dec}}^{-1})=Q(E),
	\end{equation} 

\noindent where $Q(E)$ is the injection function. The corresponding solution is

\begin{equation}
			N(E)= \Big | \frac{dE}{dt} \Big | _{\rm{loss}} ^{-1} \int_{E}^{E_{\rm{max}}} {Q(E')e^{-\tau (E,E')}dE'} ,
		\end{equation}

\noindent where
	
		\begin{equation} \label{eq:tau}
			\tau (E,E')= \int_{E}^{E'} {dE'' \Big |\frac{dE''}{dt} \Big | _{\rm{loss}}^{-1}  (t_{\rm{esc}}^{-1}+t_{\rm{dec}}^{-1})}.
		\end{equation}

\noindent The injection function for non-thermal protons and electrons is a powerlaw of the energy of the particles $Q(E)=Q_{0} E^{-\alpha}e^{-E/E_{\rm{max}}}$, as a consequence of the diffusive particle acceleration by shock waves. Following \citet{drury}, we  adopt a standard index $\alpha=2.2$. The normalization constant $Q_{0}$ can be obtained from the total power injected in relativistic protons and electrons, $L_{\rm{rel}}=L_{p}+L_{e}$. This power is assumed to be a fraction of the luminosity of the corona, $L_{\rm{rel}}=q_{\rm rel} L_{\rm{c}}$, with $q_{\rm rel}=10^{-2}$ \citep{blandford02}. The way in which energy is divided between hadrons and leptons is unknown, but different scenarios can be taken into account by setting $L_{p}=aL_{e}$. We consider models with $a=100$ (proton-dominated scenario, as for Galactic cosmic rays) and $a=1$ (equipartition between both species).

To obtain the pion distribution, we use the pion injection due to $pp$ collisions \citep{fletcher} and $p \gamma$ interactions \citep{atoyan}. The injection of muons is calculated using the expressions from \citet{lipari}. The target fields are those of the corona in the case of matter, and the corona-plus-disk for photons. Once the dominant energy loss process is identified, the transport equations can be decoupled without a significant loss of accuracy (say less than $\sim 10 \%$). This approximation simplifies considerably the calculations. A full treatment is being developed by the authors and preliminary results have been used for comparison and accuracy estimates.

\section{Spectral energy distributions}\label{SEDs}

\subsection{Radiative processes}

Expressions to calculate the synchrotron spectrum can be found, for example, in \citet{blumenthal}. The power radiated by a single particle of energy $E$ and pitch angle $\alpha$ is given by

\begin{equation}\label{eq:aprox}
			P_{\rm{synchr}}(E,E_{\gamma})=\frac{\sqrt{3}e^3B \sin \alpha}{hmc^2} \frac{E_{\gamma}}{E_{\rm{c}}} \int_{E_{\gamma}/E_{c}}^{\infty}{K_{5/3}(\xi)d\xi},
		\end{equation}
		
\noindent where $K_{5/3}(\xi)$ is a modified Bessel function and the characteristic energy is

		\begin{equation}
		E_{\rm{c}}=\frac{3}{4\pi}\frac{ehB \sin \alpha}{mc} \left(\frac{E}{mc^2} \right)^2.
		\end{equation}

\noindent For a particle distribution $N(E)$ and a volume of the emission region $V$, the total luminosity is

\begin{equation}
			L_{\gamma}(E_{\gamma})  = E_{\gamma} \int_{V_{\rm{c}}}  d^3 r  \int_{E_{\rm{min}}}^{E_{\rm{max}}} {dE N(E)P_{\rm{synchr}}(E,E_{\gamma})} .
		\end{equation}

According to \citet{blumenthal}, the spectrum of photons scattered by an electron of energy $E_{e}=\gamma _{e}m_{e}c^2$ in a target radiation field of density $n_{\rm{ph}}(\epsilon)$ is

\begin{equation}
		P_{\rm{IC}}(E_{e},E_{\gamma},\epsilon)= \frac{3\sigma_{\rm{T}} c(m_{e}c^2)^2}{4E_{e}^2} \frac{n_{\rm{ph}}(\epsilon)}{\epsilon}F(q),
		\end{equation}
		
\noindent where $F(q)$, $q$, and $\Gamma$ are given by Eq. (\ref{IC}) (in this case $\epsilon_{1}=E_{\gamma}$). The allowed range of energies for the scattered photons is 

	\begin{equation}
		\epsilon \leq E_{\gamma} \leq \frac{\Gamma}{1+\Gamma}E_{e} \textrm{ .}
	\end{equation}

\noindent The total luminosity can then be obtained from

	\begin{eqnarray}
	L_{\rm{IC}}(E_{\gamma}) = E_{\gamma}^2 \int_{V_{\rm{c}}} d^3 r \int_{E_{\rm{min}}}^{E_{\rm{max}}}& dE_{e} &  N_{e}(E_{e}) \nonumber\\
	&& \times  \int _{\epsilon_{\rm{min}}}^{\epsilon_{\rm{max}}} d\epsilon P_{\rm{IC}} .
	\end{eqnarray}
	
\noindent This equation takes into account the Klein-Nishina effect on the cross-section with energy.

The most important products of $p \gamma$ interactions at high energies are pions. Neutral $\pi$-mesons decay with a probability of 98.8 \% into two gamma rays

	\begin{equation}
		\pi^{0} \rightarrow \gamma + \gamma.
	\end{equation}
	
\noindent To estimate the spectrum from the decay of $\pi^{0}$, the $\delta$-functional formalism can be applied \citep{atoyan}. The emissivity of neutral pions in this approximation is

	\begin{equation}
		Q_{\pi^0}^{(p\gamma)}(E_{\pi}) = 5N_{p}(5E_{\pi})\omega_{p\gamma}(5E_{\pi})n_{\pi^0}(5E_{\pi}) \textrm{ ,}
	\end{equation}

\noindent where $\omega_{p\gamma}$ is the collision rate given by

\begin{equation}
			\omega_{p\gamma}(E_{p})= \frac{m_{p}^2c^5}{2E_{p}^2}\int^{{\infty}}_{\frac{\epsilon_{\rm{th}}}{2\gamma_{p}}} 				{d\epsilon \frac{n_{\rm{ph}}(\epsilon)}{\epsilon^2}} \int^{2 \epsilon	E_{p}\gamma_{p}}_{\epsilon_{\rm{th}}}{d\epsilon '\sigma_{p\gamma}^{(\pi)}(\epsilon ')\epsilon '} ,
	\end{equation}
	
\noindent and $n_{\pi^0}$ is the mean number of neutral pions created per collision. For more details, we refer to \citet{atoyan}.

Taking into account that each $\pi^{0}$ decays into two photons, the photon emissivity is given by	

\begin{eqnarray}
		q_{\gamma}(E_{\gamma}) & = &2 \int {Q_{\pi^0}^{(p\gamma)}(E_{\pi}) \delta(E_{\gamma}-0.5E_{\pi}) dE_{\pi}} \nonumber\\
	  & = & 20N_{p}(10E_{\gamma})\omega_{p\gamma ,\pi}(10E_{\gamma})n_{\pi^0}(10E_{\gamma}) \textrm{ .}
	\end{eqnarray}
	
\noindent It is possible to estimate the injection function of pions produced by $pp$ interactions also using the $\delta$-functional formalism \citep{aha03}. In this approximation, the emissivity of $\pi^0$ is
	
	\begin{eqnarray}
		Q_{\pi^0}^{(pp)}(E_{\pi}) & = & cn_{i} \int \delta(E_{\pi}-\kappa_{\pi}E_{\rm{kin}}) \sigma_{pp}(E_{p})N_{p}(E_{p})dE_{p} \nonumber\\
		      & = & \frac{cn_{i}}{\kappa_{\pi}} \sigma_{pp} \left(m_{p}c^2+\frac{E_{\pi}}{\kappa_{\pi}} \right) N_{p} \left(m_{p}c^2+\frac{E_{\pi}}{\kappa_{\pi}} \right) \textrm{.}
	\end{eqnarray}
 \noindent For proton energies in the GeV-TeV range, $\kappa \approx 0.17$ \citep{gaisser} and the total cross-section $\sigma_{pp}$ can be approximated by Eq. (\ref{sigma_pp}) for $E_{\rm{kin}} \geq 1$ GeV, and $\sigma_{pp} \approx 0$ for $E_{\rm{kin}} < 1$ GeV.	
	
		Once the pion injection $Q_{\pi^0}(E_{\pi})$ is known, the gamma-ray emissivity $q_{\gamma}(E_{\gamma})$ (erg$^{-1}$ cm$^{-3}$ s$^{-1}$) can be obtained from
	
	\begin{equation}
	q_{\gamma}(E_{\gamma})=2 \int^{\infty}_{E_{\rm{min}}} \frac{Q_{\pi^0}^{(pp)}(E_{\pi})}{\sqrt{E_{\pi}^2-m_{\pi}^2c^4}} dE_{\pi} \textrm{,}
	\end{equation}
	
\noindent where $E_{\rm{min}}=E_{\gamma}+m_{\pi}^2c^4/4E_{\gamma}$.

\subsection{Secondary pair injection}

Two channels for secondary pair production were taken into account: pairs produced in photon-photon interactions and electron and positron production as a result of muon decay.

The energy spectrum of pairs produced in photon-photon interactions was studied, for example, by \citet{aha01}. Under the conditions $\epsilon \ll m_{e}c^2 \leq E_{\gamma}$, the pair injection $Q_{e}(\gamma_{e})$ (in units of erg $^{-1}$ s$^{-1}$ cm$^{-3}$) can be approximated by the expression

	\begin{eqnarray}
		Q_{e}(E_{e}) & = & \frac{3}{32} \frac{c\sigma_{\rm{T}}}{m_{e}c^2} \int\limits^{\infty}_{\gamma_{e}} \int\limits^{\infty} _{\frac{\epsilon_{\gamma}}{4\gamma_{e}(\epsilon_{\gamma}-\gamma_{e})}} d\epsilon_{\gamma} d\omega 
		\frac{n_{\gamma}(\epsilon_{\gamma})}{\epsilon_{\gamma}^3} \frac{n_{\rm{ph}}(\omega)}{\omega^2} \nonumber \\
		& \times & \Biggl\{ \frac{4\epsilon_{\gamma}^2}{\gamma_{e}(\epsilon_{\gamma}-\gamma_{e})}  
		 \ln \Big [ \frac{4\gamma_{e}\omega(\epsilon_{\gamma}-\gamma_{e})}{\epsilon_{\gamma}} \Big ] -8\epsilon_{\gamma}\omega +  \nonumber \\
		 & + & \frac{2(2\epsilon_{\gamma}\omega-1)\epsilon_{\gamma}^2}{\gamma_{e}(\epsilon_{\gamma}-\gamma_{e})} - \left ( 1- \frac{1}{\epsilon_{\gamma}\omega} \right ) \frac{\epsilon_{\gamma}^4}{\gamma_{e}^2(\epsilon_{\gamma}-\gamma_{e})^2} \Biggr\} ,
	\end{eqnarray}

\noindent where $\gamma_{e}=E_{e}/m_{e}c^2$ is the Lorentz factor of the electron, $\epsilon_{\gamma}=E_{\gamma}/m_{e}c^2$, and $\omega=\epsilon/m_{e}c^2$ are the dimensionless photon energies.

The second production mechanism of secondary electrons/positrons is the decay of charged pions, given by

\begin{equation}
		 \pi^{\pm} \rightarrow \mu^{\pm} + \nu_{\mu}(\overline{\nu}_{\mu}) \textrm{,}
		\end{equation}
	
	\begin{equation}
		 \mu^{\pm} \rightarrow e^{\pm} + \overline{\nu}_{\mu}(\nu_{\mu}) + \nu_{e}(\overline{\nu}_{e}) .
		\end{equation}
		
 \noindent This process has been extensively studied by several authors. Given a muon distribution $N_{\mu}(\gamma_{\mu})$, it is possible to obtain the pair injection $Q_{e}(\gamma_{e})$ as a result of this process using \citep{ramaty}

\begin{equation}
	Q_{e}(\gamma_{e})= \int\limits_{1}^{\gamma_{e}'^{\rm{max}}} d\gamma_{e}' \frac{1}{2} \frac{P(\gamma_{e}')}{\sqrt{\gamma_{e}'^2-1}} \int\limits_{\gamma_{\mu}^{-}}^{\gamma_{\mu}^{+}} d\gamma_{\mu} \frac{N_{\mu}(\gamma_{\mu}) t^{-1}_{\rm{dec}}(\gamma_{\mu})}{\sqrt{\gamma_{\mu}^2-1}} ,
\end{equation}
  
  \noindent where $\gamma_{e}'^{\rm{max}}=104$, 
  
\begin{equation}
	\gamma_{\mu}^{\pm}=\gamma_{e}\gamma_{e}' \pm \sqrt{\gamma_{e}^2-1}\sqrt{\gamma_{e}'^2-1},
\end{equation}

\noindent and the electron distribution in the rest-frame of muon is

\begin{equation}
	P(\gamma_{e}')= \frac{2\gamma_{e}'^2}{\gamma_{e}'^{\rm{max}\;3}} \Big[ 3- \frac{2\gamma_{e}'}{\gamma_{e}'^{\rm{max}}} \Big].
\end{equation}

\subsection{Spectral energy distributions}

\begin{figure*}[!t]
\centering
\subfigure[Case $a=1$, diffusion.]{\label{fig:SEDs:a}\includegraphics[width=0.45\textwidth, keepaspectratio]{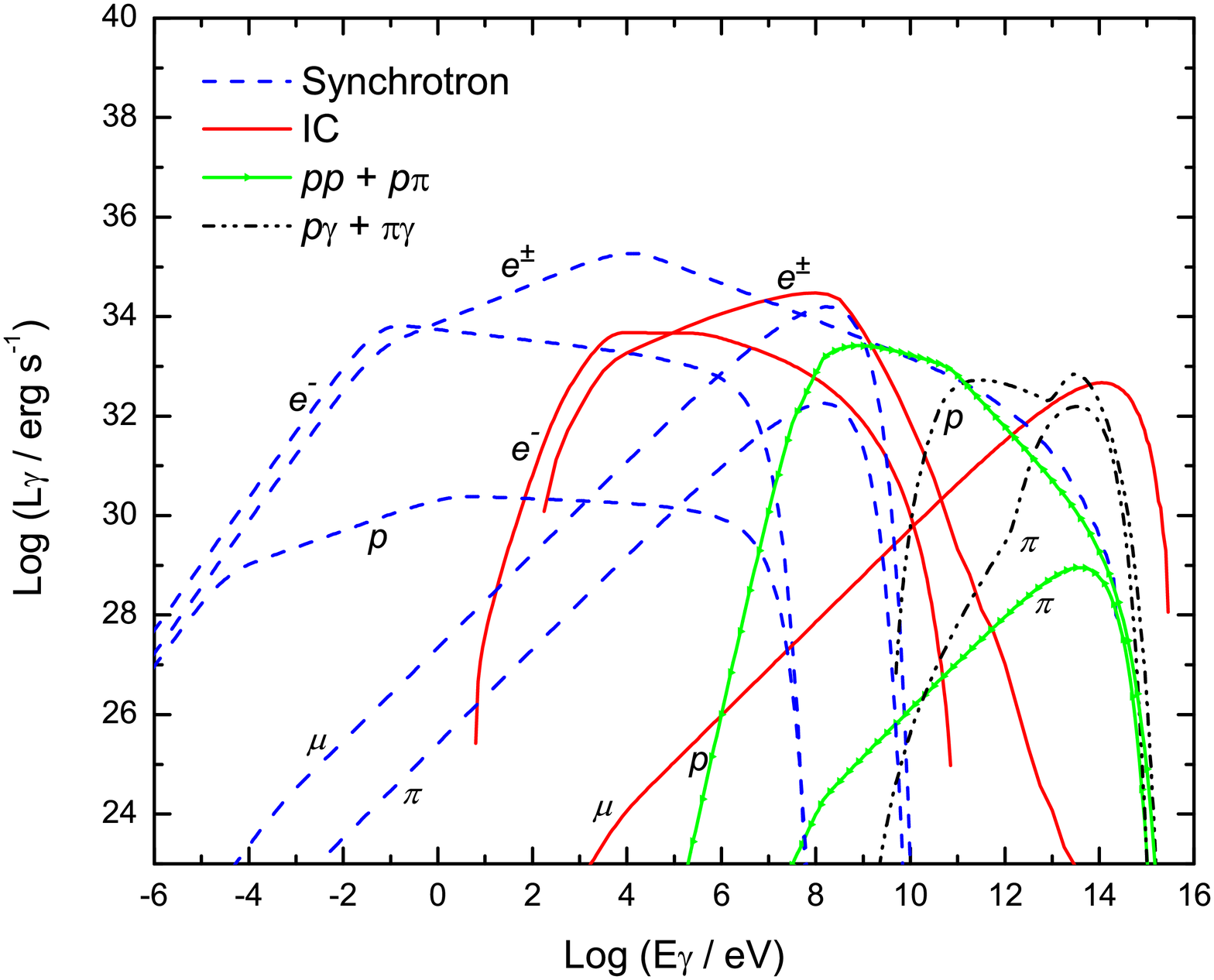}} \hspace{20pt}
\subfigure[Case $a=1$, advection.]{\label{fig:SEDs:b}\includegraphics[width=0.45\textwidth, keepaspectratio]{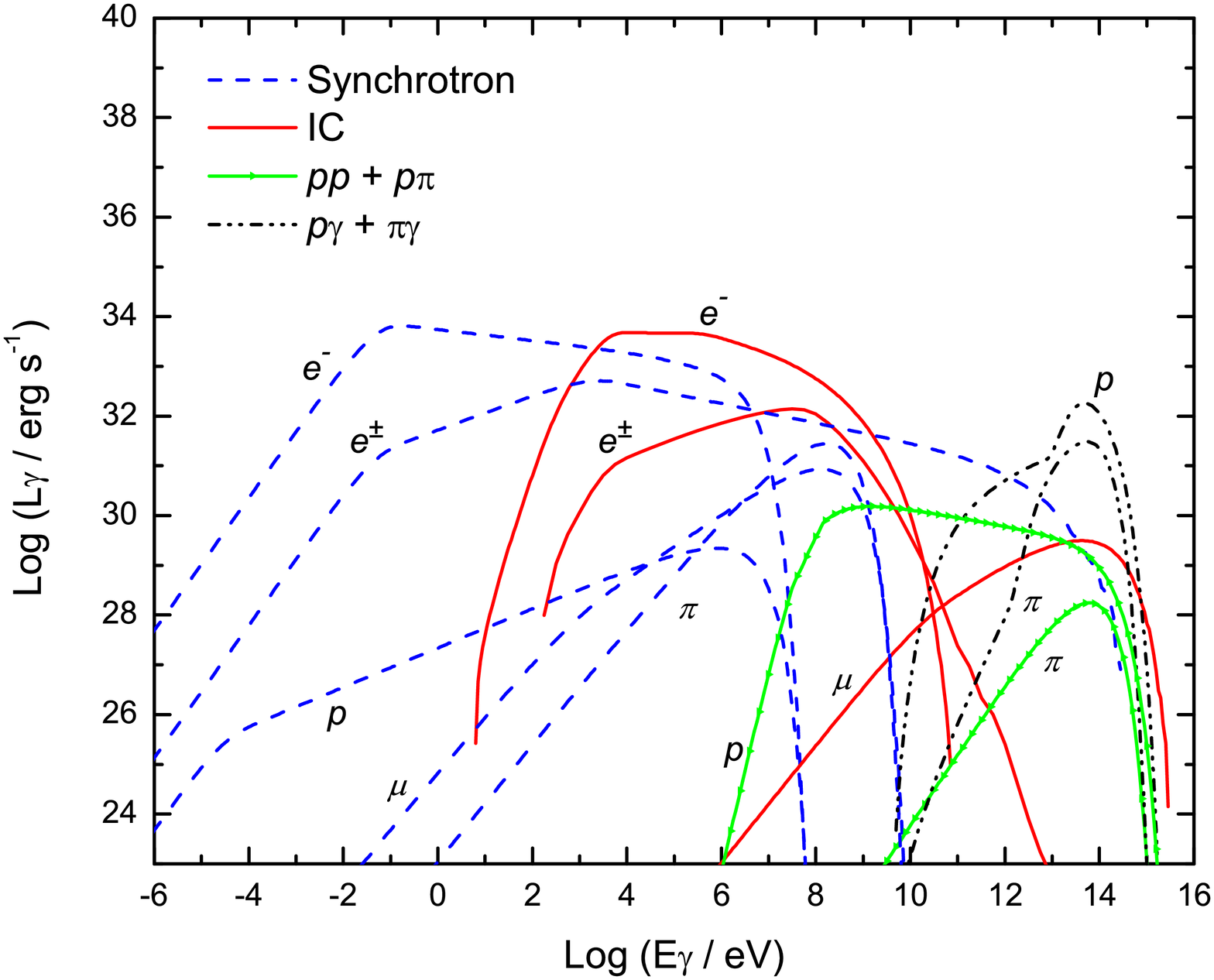}} \hfill \\
\subfigure[Case $a=100$, diffusion.]{\label{fig:SEDs:c}\includegraphics[width=0.45\textwidth, keepaspectratio]{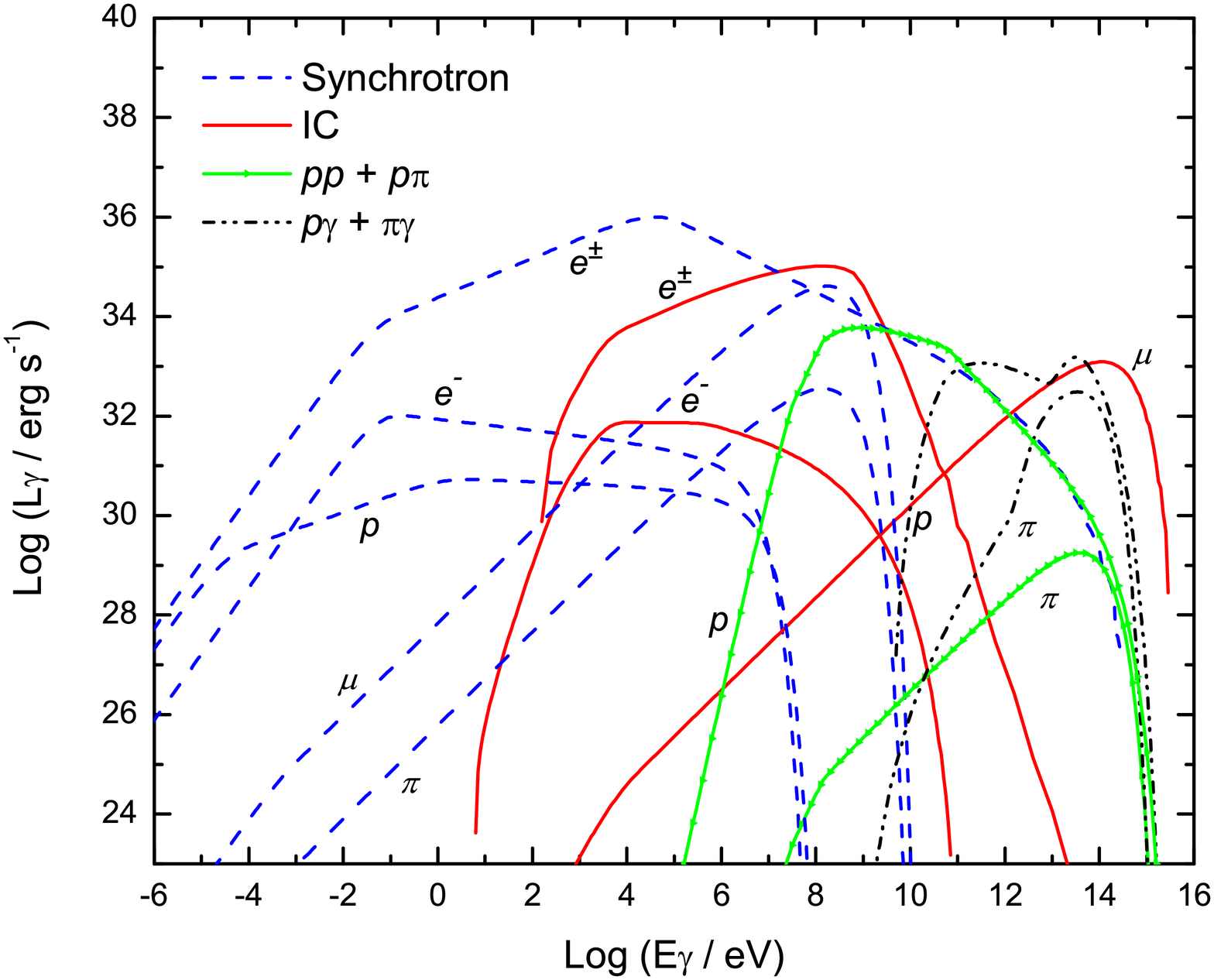}} \hspace{20pt}
\subfigure[Case $a=100$, advection.]{\label{fig:SEDs:d}\includegraphics[width=0.45\textwidth, keepaspectratio]{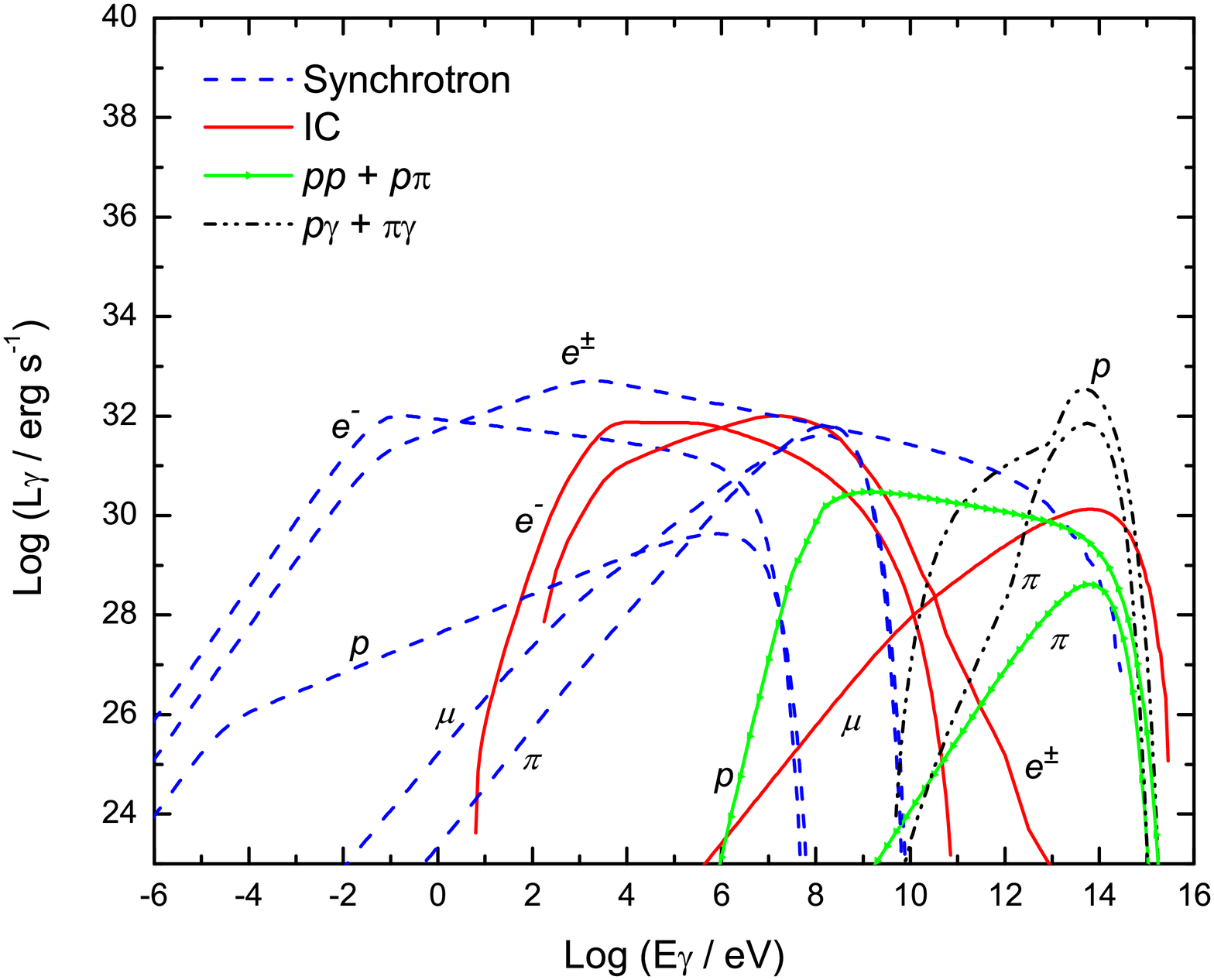}} \hfill 
\caption{Spectral energy distributions obtained for different sets of parameters. Internal absorption not included.}
\label{fig:SEDs}
\end{figure*}

Figure \ref{fig:SEDs} shows all contributions to the total luminosity for different advection regimes and for different values of the parameter $a$. On the one hand, it can be seen that the luminosities produced by primary electrons are higher in models with $a=1$. On the other hand, luminosities produced by hadrons and muons are higher in models with a static corona and diffusion of the relativistic particles. This is because in models with advection an important fraction of protons and pions are swallowed by the black hole, while with diffusion these particles are able to lose their energy before falling onto the compact object or escaping from the system. In models with $a=100$, the non-thermal emission at $E_{\gamma}>1$ MeV is dominated by synchrotron and IC radiation of secondary pairs. At very high energies, the main contributions to the spectrum are due to photo-meson production in all models. We note that below $\sim 150$ keV the source will be totally dominated by thermal Comptonization (not shown in the figures for clarity).

From Fig. \ref{fig:SEDs}, we conclude that there are two parameters that determine the relevant radiative processes: the hadronic content in the plasma and the advection velocity. If the hadronic content is high, then a large number of secondary particles are expected to increase the emission at high energies. This is precisely what happens in our model for a high value of the parameter $a$. However, the advection also has an important role, because in a corona with convection a significant part of the proton content will be engulfed by the black hole reducing the emission. The overall SED predicted by a particular model is then the result of the specific balance between the two main free parameters.

There are other physical quantities that are important to our model, such as the magnetic field and the background thermal luminosity, but the values of these parameters are restricted by observations of sources, such as Cygnus X-1.

\subsection{Absorption}

\begin{figure*}[!t]
\centering
\subfigure[Case $a=1$, diffusion.]{\label{fig:Ltotal:a}\includegraphics[width=0.45\textwidth, keepaspectratio]{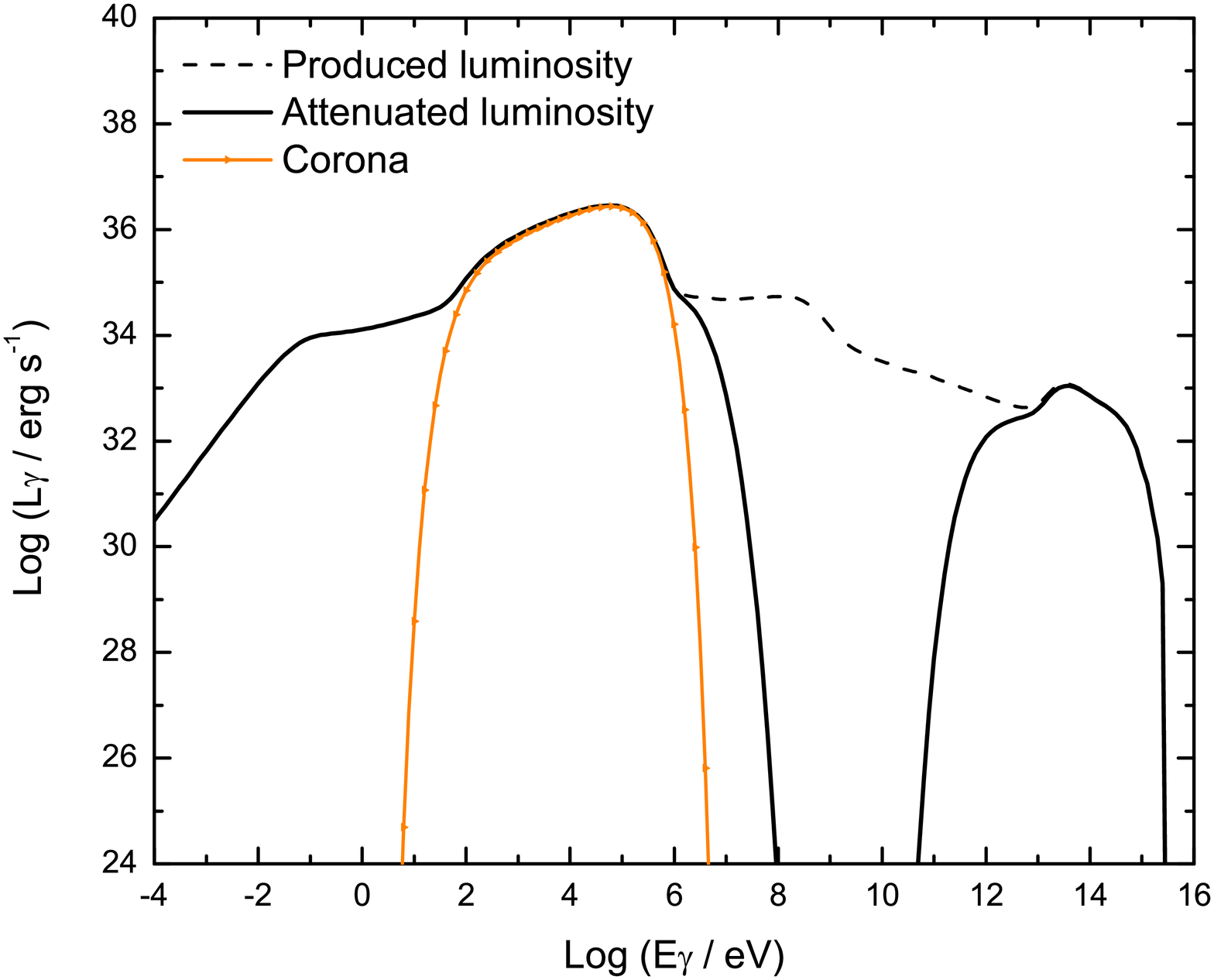}} \hspace{20pt}
\subfigure[Case $a=1$, advection.]{\label{fig:Ltotal:b}\includegraphics[width=0.45\textwidth, keepaspectratio]{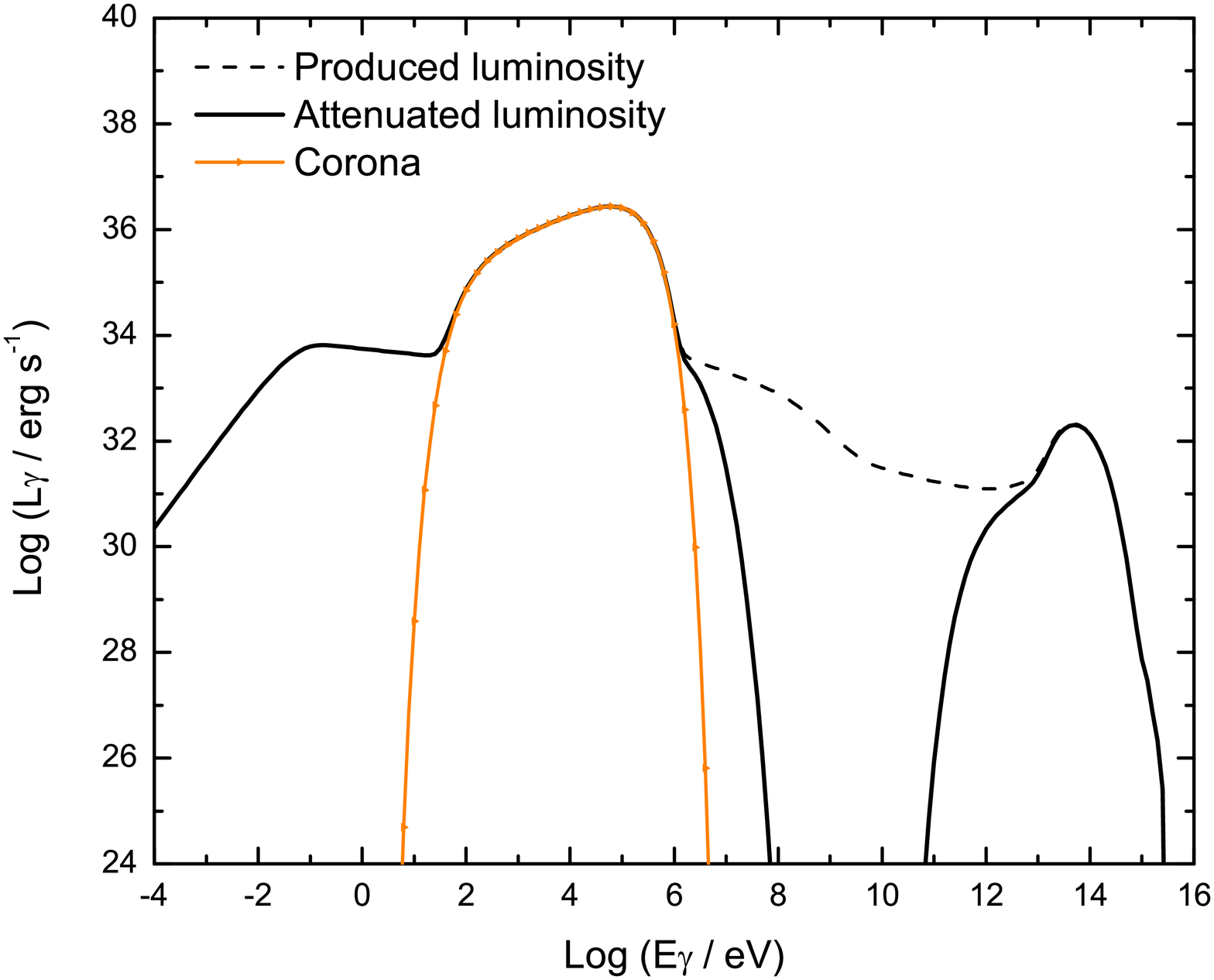}}  \hfill \\
\subfigure[Case $a=100$, diffusion.]{\label{fig:Ltotal:c}\includegraphics[width=0.45\textwidth, keepaspectratio]{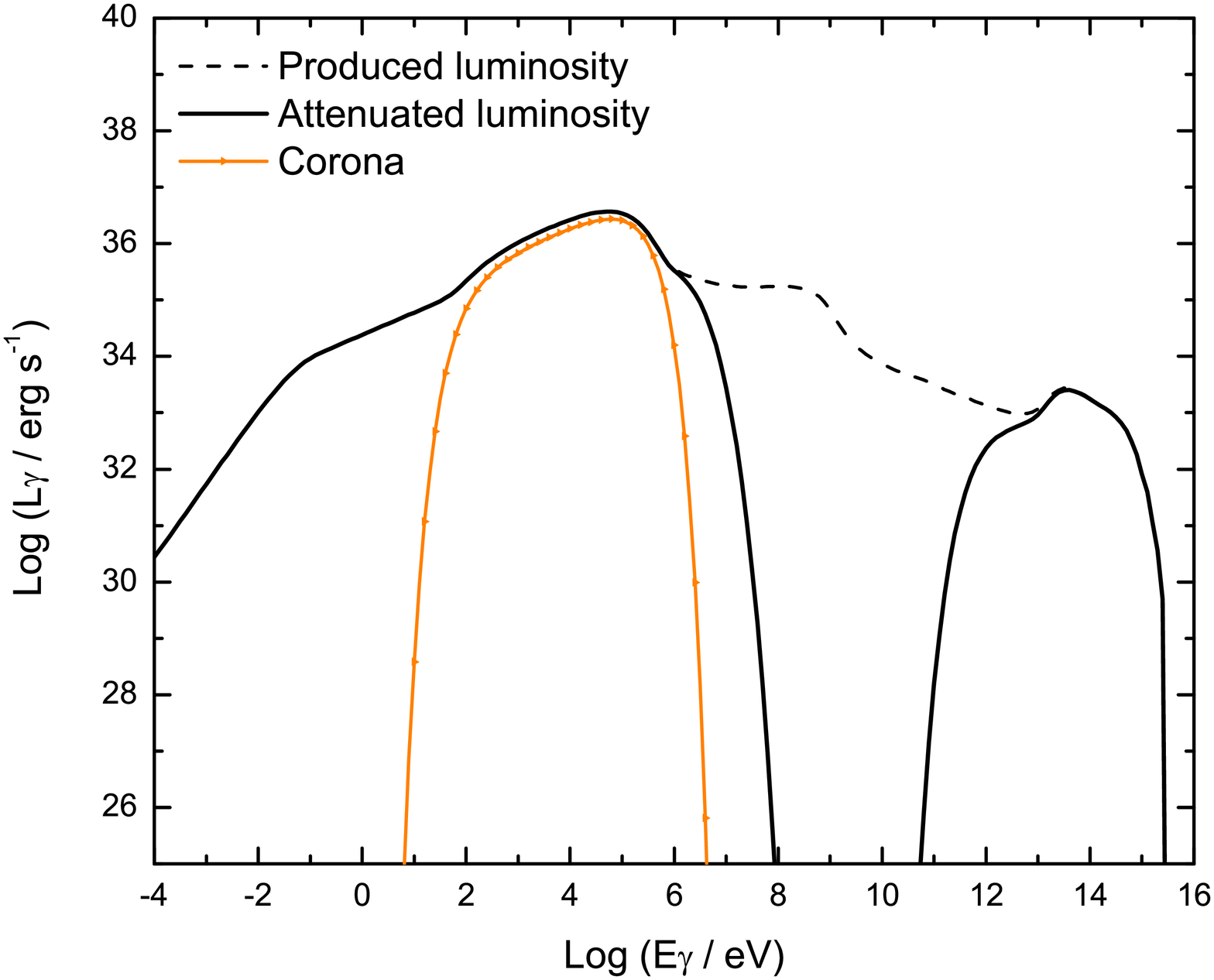}} \hspace{20pt}
\subfigure[Case $a=100$, advection.]{\label{fig:Ltotal:d}\includegraphics[width=0.45\textwidth, keepaspectratio]{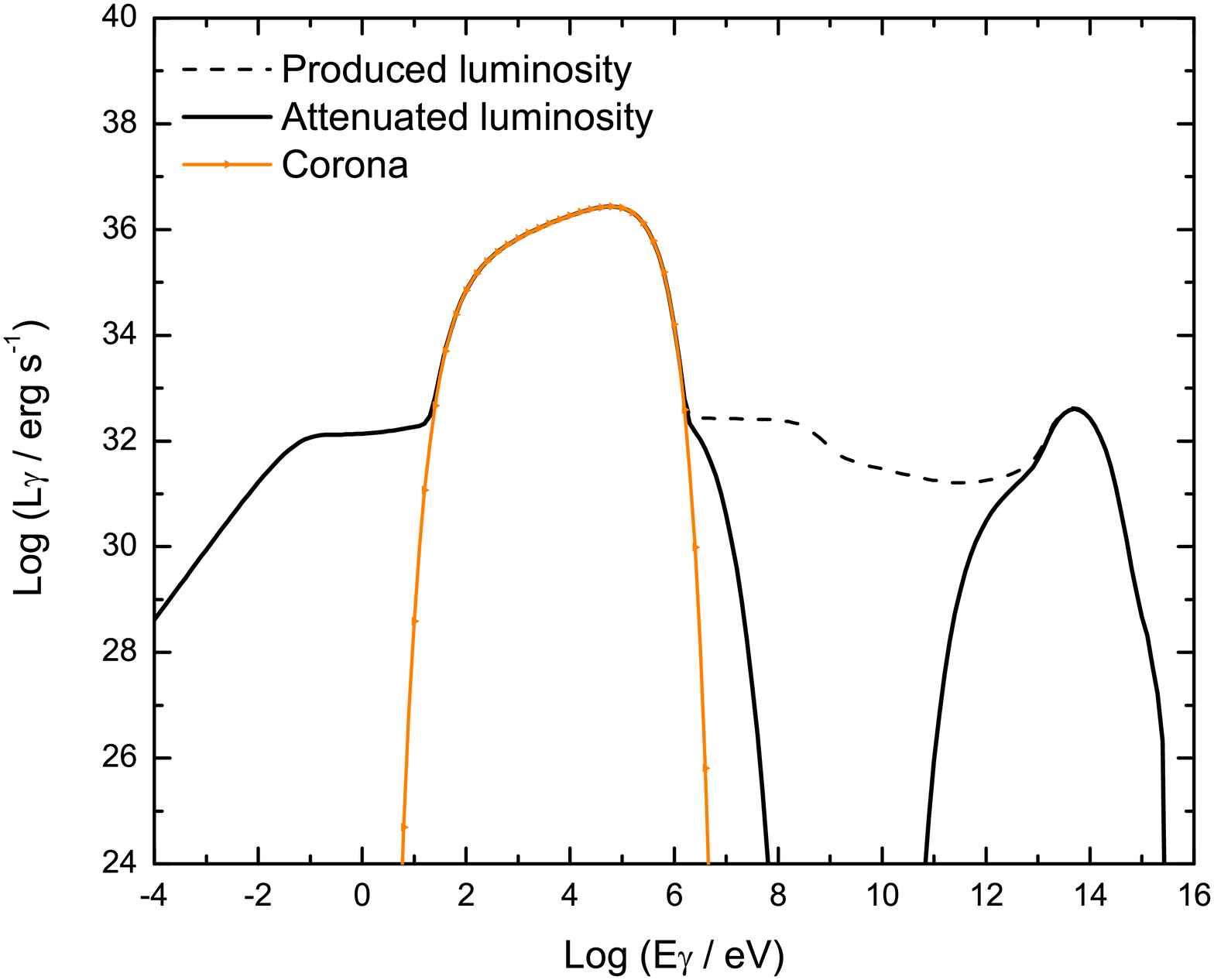}}  \hfill
\caption{Produced luminosity and attenuated luminosity for different sets of parameters.}
\label{fig:Ltotal}
\end{figure*}

Gamma-rays created in astrophysical sources can be absorbed by different mechanisms. The absorption can be quantified by the absorption coefficient or opacity $\tau$. If the original gamma-ray luminosity is $L_{\gamma}^0(E_{\gamma})$, the attenuated luminosity $L_{\gamma}(E_{\gamma})$ after the photon travels a distance $l$ is

	\begin{equation}
		L_{\gamma}(E_{\gamma})=L_{\gamma}^0(E_{\gamma}) e^{-\tau(l,E_{\gamma})}.
	\end{equation}
	
	\noindent The opacity depends on the absorption process. In the model presented here, the main mechanism of absorption is photon-photon pair production. This interaction takes place between gamma-rays and thermal X-rays photons from the corona. This process is possible only above a kinematic energy threshold given by
	
	\begin{equation}
		E_{\gamma}\epsilon > (m_{e}c^2)^2,
	\end{equation}
	
\noindent in a frontal collision.	
	
	For a photon created at a distance $r$ from the center of the corona, the opacity caused by photon-photon pair production can be obtained from \citep{romero03}
	
	\begin{equation}
		\tau_{\gamma\gamma}(E_{\gamma},r)= \int_{E_{\rm{min}}}^{\infty} \int_{r}^{\infty} n_{\rm{ph}}(\epsilon,r') \sigma_{\gamma\gamma}(\epsilon,E_{\gamma}) dr' d\epsilon ,
	\end{equation}
	
	\noindent where $\epsilon$ is the energy of the X-ray photons and $n_{\rm{ph}}$ their density. The total cross-section is given by the expression \citep[e.g.,][]{aha02}

	\begin{equation}
		\sigma_{\gamma\gamma}= \frac{\pi r_{0}^2}{2} (1-\beta^2) \Big[ (3-\beta^4) \ln \left( \frac{1+\beta}{1-\beta} \right) +2\beta(\beta^2-2) \Big] ,
	\end{equation}
	
\noindent where 

	\begin{equation}
		\beta = \left( 1-\frac{(m_{e}c^2)^2}{E_{\gamma}\epsilon} \right)^{1/2}.   
	\end{equation}
	
	Figure \ref{fig:Ltotal} shows the difference between the produced luminosity and the attenuated luminosity. For 10 MeV $<E_{\gamma}<$ 1 TeV, almost all photons are absorbed. 
	
	Although the gamma-ray distribution depends on the electron/positron distributions and viceversa, we treat each population independently. This assumption is justified because the strong magnetic field is understood to prevent the development of cascades \citep{pellizza}. 
	
The main cooling channel for high-energy electrons/positrons is synchrotron radiation, since the IC interaction occurs mainly in the Klein-Nishina regime. This can be clearly seen from the first panel of Fig. \ref{fig:perdidas}. The synchrotron cooling then halts the development of high-energy cascades in the corona \citep[see][]{bosch01}. Only at low energies ($E_{\gamma}<100$ MeV) may a third generation of pairs be produced \citep{vieyro01}.

\section{Application to Cygnus X-1}\label{Cyg}

We applied the model discussed in the previous sections to the well-known binary system Cygnus X-1. Cygnus X-1 is a very bright X-ray binary consisting of a compact object of $\sim 10.1$ $M_{\odot}$ and a companion O9.7 Iab star of $\sim17.8$ $M_{\odot}$ \citep{herrero}, at an estimated distance of $\sim 2$ kpc (e.g., \citealt{gierl} and references therein). The X-ray emission alternates between soft and hard states. The spectrum in both states can be approximately represented as the sum of a black body and a powerlaw with exponential cutoff \citep[e.g.,][]{poutanen01}. During the soft state, the black body component is dominant and the powerlaw is steep, with a photon spectral index $\sim 2.8$ \citep[e.g.,][]{frontera}. During the hard state, more energy is in the powerlaw component, which is even hardened, with photon indices $\sim 1.6$ \citep[e.g.,][]{gierl}.  

\citet{McConnell} reported a high-energy tail in the low X-ray state of the source, extending from 50 keV to $\sim 5$ MeV. The data at MeV energies, collected with the COMPTEL instrument of the Compton Gamma-Ray Observatory,  can be described as a powerlaw with a photon spectral index of $3.2$. Observations with the INTEGRAL satellite have confirmed the existence of a supra-thermal tail in the spectrum \citep{cadolle}. So-called hybrid thermal/non-thermal models from \citet{poutanen02} and \citet{coppi} have been used to fit the observed spectrum \citep{McConnell,cadolle}. These models consider a hybrid pair plasma with a non-thermal component. In particular, using the EQPAIR code, \citet{McConnell} concluded that either the magnetic field in Cyg X-1 is substantially below equipartition (at least to within an order of magnitude) or the observed photon tail has a different origin than that related to locally accelerated electrons. 

\begin{figure*}[!t]
\centering
\includegraphics[width=12 cm]{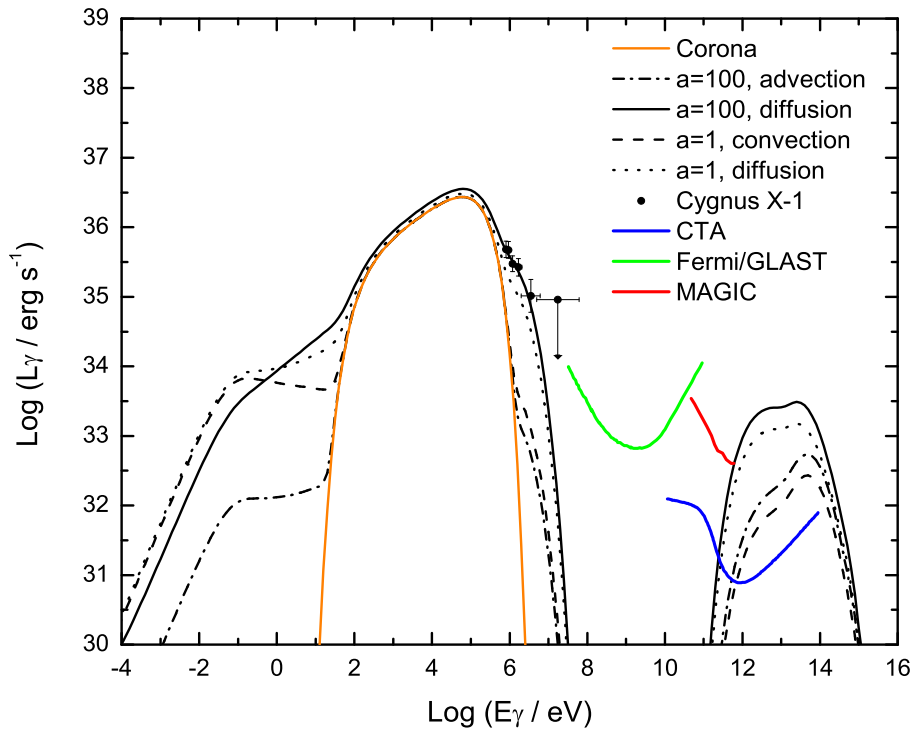}
\caption{Application of the model to Cygnus X-1. Observational data from COMPTEL \citep{McConnell}. We include the sensitivity of MAGIC, Fermi, and CTA. }
\label{fig:TODAS/Cyg}  
\end{figure*}

We applied our lepto-hadronic model to Cygnus X-1. In Fig. \ref{fig:TODAS/Cyg}, we show the predictions of the models discussed in the previous sections (equipartition and proton dominance) to Cygnus X-1. All our models assume equipartition magnetic fields. As expected, the emission in the MeV range is dominated by products of hadronic interactions and secondary pairs. The best fits are for a model with $a=100$ and little or null advection, with absorption playing a major role in shaping the spectrum. At high-energies $E_{\gamma}>1$ TeV, a bump produced mainly by photo-meson production appears. It might be easily detectable by the future Cherenkov Telescope Array, if Cygnus X-1 is within its declination range. In Fig. \ref{fig:TODAS/Cyg} we indicate the sensitivity of different instruments, including MAGIC, Fermi, and CTA. 

MAGIC detected a high-energy flare from Cygnus X-1 \citep{albert}, which, however, was likely produced in the jet of the object \citep[e.g.,][]{bosch02}. Similar considerations apply to the flare detected by the AGILE satellite \citep{sabatini}. In our model, even if large magnetic reconnection events were to modify the non-thermal population on short timescales, the GeV emission would be totally suppressed by photon annihilation in the thermal bath of the corona. Gamma-ray flaring events at GeV energies, then, cannot arise from a strongly magnetized corona.

\section{Discussion}\label{Discussion}

Proton-proton interactions in a purely thermal non-magnetized corona were considered by \citet{eilek} and \citet{mahadevan}. The latter authors also considered the effects of a non-thermal population of protons, but without taking into account pair production in the soft photon field, which was considered by \citet{jourdain} and independently by \citet{bhattacharyya01,bhattacharyya02}. The effects of the magnetic field were not considered by any of these authors. 

The role of the magnetic field in the cooling of the different types of relativistic particles in the corona is very important and cannot be ignored, as it can be clearly seen in Fig. \ref{fig:perdidas}. Electrons are cooled almost immediately by synchrotron losses in fields of $\sim 10^5$ G, truncating the development of IC cascades. Photo-meson and photo-pair production in models with large values of the parameter $a$ can produce significant high-energy gamma-ray emission and secondary pair injection. The high-energy emission may be detectable in the future from different sources. Since it is of hadronic origin, detections or upper limits can be used to place constraints on the number of relativistic protons in the corona. 

Emission in the range 100 MeV to 1 TeV is not expected because of absorption in the soft photon field. All emission detected in this range should be produced in the jet \citep[e.g.,][]{romero01,romero02,bosch02}. However, if a sudden injection of relativistic protons occurs, for instance as a consequence of major reconnection events, a neutrino burst may be produced. This possibility will be explored in detail in a forthcoming paper \citep{vieyro02}.

\section{Conclusions}\label{Conclusions}

We have developed a model to predict the radiative output of a two-temperature magnetized corona with a mixed population of non-thermal electrons and protons. All radiative processes have been calculated in this scenario. The contribution of secondary pairs and transient particles such as muons and charged pions has been shown to be important. The complexity of the spectral energy distributions calculated reflects the variety of the physical processes involved. We have applied our model to the case of Cygnus X-1 and obtained a good agreement with the observed soft gamma-ray tail observed by COMPTEL and INTEGRAL. We have also predicted a high-energy bump at $E>1$ TeV that might be detectable with future giant telescope arrays. Additional research into the inhomogeneous coronae is in progress.   

\section*{Acknowledgments}

We thank an anonymous referee for insightful comments and suggestions. This research was supported by ANPCyT through grant PICT-2007-00848 BID 1728/OC-AR and by the Ministerio de Educaci\'on y Ciencia (Spain) under grant AYA 2007-68034-C03-01, FEDER funds.

\appendix

\section{Bethe-Heitler process and relevant approximations}

The cross-section for proton-photon pair production is also known as the Bethe-Heitler cross-section. This function increases monotonically with photon energy. In the limits of low and high energy, the analytical approximations are given by \citep{begelman}
	
			\begin{equation}
				\sigma^{(e)}(x')\approx 1.2\times10^{-27}\left(\frac{x'-{x'}_{th}^{(e)}}{{x'}_{th}^{(e)}}\right)^3 \textrm{cm}^2 
			\end{equation}	
for $2 \geq x'\leq 4 $, and

		\begin{eqnarray}
				\sigma^{(e)}(x') &  \approx & 5.8\times10^{-28} \Big[ 3.1\ln(2x')-8.1+ \nonumber \\ 
				& +& \left(\frac{2}{x'}\right)^2 \Big(2.7\ln(2x') -\ln^2(2x')  \\
				&+& 0.7\ln^3(2x')+0.5 \Big) \Big]  \textrm{cm}^2 \nonumber \textrm{,}
		\end{eqnarray}

\noindent for $x'\geq 4$. The variable $x'$ is such that $\epsilon' ={x'}m_{e}c^2$ is the photon energy measured in the proton rest-frame and ${x'}_{\rm{th}}^{(e)}=2$ is the threshold energy. The inelasticity of this process has a maximum at  ${x'}_{\rm{th}}^{(e)}$, and decreases monotonically with photon energy. The low and high energy approximations are
	
	\begin{eqnarray}
		K^{(e)}(x') & = & 4\frac{m_e}{m_p}x'^{-1} \Big[1+0.4 \ln{(x'-1)}+0.1{\ln{(x'-1)}}^2+  \nonumber\\
		&+& 0.0078{\ln{(x'-1)}}^3 \Big]  
	\end{eqnarray}
for $x' \leq 1000$, and

		\begin{eqnarray}
		K^{(e)}(x') &=& 4\frac{m_e}{m_p}x'^{-1} \nonumber \\
		 & \times & \left( \frac{-8.8+5.6\ln{x'}-1.6{\ln{x'}}^2 +0.7{\ln{x'}}^3} {3.1{\ln{2x'}}-8.1}\right) 
		\end{eqnarray}
for $x'\geq 1000$.

The resulting pairs are then considered as a secondary source of leptons when estimating the spectral energy distribution. This contribution is, in general, negligible compared to other sources of secondary pairs in the context discussed in this paper.

\bibliography{bibliografia}
\bibliographystyle{aa}


\end{document}